\documentstyle[12pt]{article}
\def\ie{{\em i.e.}}
\def\eg{{\em e.g.}}

\def\beq{\begin{equation}}
\def\eeq{\end{equation}}

\catcode`\@=11 % This allows us to modify PLAIN macros.
\def\coeff#1#2{{\textstyle{#1\over #2}}}

\def\lsim{\mathrel{\mathpalette\@versim<}}
\def\gsim{\mathrel{\mathpalette\@versim>}}
\def\@versim#1#2{\vcenter{\offinterlineskip
    \ialign{$\m@th#1\hfil##\hfil$\crcr#2\crcr\sim\crcr } }}
\def\etal{{\em et. al.}}
\def\JL{J. L. Lopez}
\def\DVN{D. V. Nanopoulos}
\def\AZ{A. Zichichi}

\def\t1{{\tilde 1}}

\def\GeV{\,{\rm GeV}}
\def\TeV{\,{\rm TeV}}

\def\to{\rightarrow}
\def\pb{\,{\rm pb}}
\def\ipb{\,{\rm pb}^{-1}}
\def \met	{/\!\!\!\!E_{T}}
\def\NPB#1#2#3{Nucl. Phys. B {\bf#1} (19#2) #3}
\def\PLB#1#2#3{Phys. Lett. B {\bf#1} (19#2) #3}

\def\PRD#1#2#3{Phys. Rev. D {\bf#1} (19#2) #3}
\def\PRL#1#2#3{Phys. Rev. Lett. {\bf#1} (19#2) #3}
\def\PRT#1#2#3{Phys. Rep. {\bf#1} (19#2) #3}
\def\MODA#1#2#3{Mod. Phys. Lett. A {\bf#1} (19#2) #3}
\def\IJMP#1#2#3{Int. J. Mod. Phys. A {\bf#1} (19#2) #3}
\def\TAMU#1{Texas A \& M University preprint CTP-TAMU-#1}

\begin{document}
% TH format
\begin{flushright}
\baselineskip=12pt
{CERN-TH.7296/94}\\
{CTP-TAMU-31/94}\\
{ACT-10/94}\\
%{\bf Preliminary version}
\end{flushright}

\begin{center}
\vglue 0.5cm
{\Huge\bf A light stop and its consequences at the Tevatron and
LEPII\\}
\vglue 1cm
{JORGE L. LOPEZ$^{(a),(b)}$, D. V. NANOPOULOS$^{(a),(b),(c)}$, and A.
ZICHICHI$^{(d)}$\\}
\vglue 0.4cm
{\em $^{(a)}$Center for Theoretical Physics, Department of Physics, Texas A\&M
University\\}
{\em College Station, TX 77843--4242, USA\\}
{\em $^{(b)}$Astroparticle Physics Group, Houston Advanced Research Center
(HARC)\\}
{\em The Mitchell Campus, The Woodlands, TX 77381, USA\\}
{\em $^{(c)}$CERN Theory Division, 1211 Geneva 23, Switzerland\\}
{\em $^{(d)}$CERN, 1211 Geneva 23, Switzerland\\}
\baselineskip=12pt

%\vglue 0.5cm
{\tenrm ABSTRACT}
\end{center}
\vglue 0.5cm
{\rightskip=3pc
 \leftskip=3pc
\noindent
\large
\baselineskip=20pt
An interesting prediction of a string-inspired {\em one-parameter} $SU(5)\times
U(1)$ supergravity model, is the fact that the lightest member ($\tilde t_1$)
of the top-squark doublet $(\tilde t_1,\tilde t_2)$, may be substantially
lighter than the top quark. This sparticle ($\tilde t_1$) may be readily
pair-produced at the Tevatron and, if $m_{\tilde t_1}\lsim130\GeV$, even be
observed at the end of Run IB. Top-squark production may also be an important
source of sought-for top-quark signatures in the dilepton and $\ell$+jets
channels. Therefore, a re-analysis of the top data
sample in the presence of a possibly light top-squark appears necessary
before definitive statements concerning the discovery of the top quark can be
made. Such a light top-squark is linked with a light supersymmetric spectrum,
which can certainly be searched for at the Tevatron through trilepton and
squark-gluino searches, and at LEPII through direct $\tilde t_1$
pair-production (for $m_{\tilde t_1}\lsim100\GeV$) and via chargino and
Higgs-boson searches.}
\vspace{1cm}

%\vspace{0.5cm}
% TH format
\begin{flushleft}
\baselineskip=12pt
{CERN-TH.7296/94}\\
{CTP-TAMU-31/94}\\
{ACT-10/94}\\
June 1994
\end{flushleft}

\vfill\eject
\setcounter{page}{1}
\pagestyle{plain}
\baselineskip=14pt

The CDF collaboration has recently announced ``evidence" for the existence
of the top quark with mass $m_t=174\pm17\GeV$ \cite{CDF}. There exists
also plenty of indirect evidence for the top quark from precise electroweak
measurements at LEP \cite{LEPmt}, when contrasted with the corresponding
theoretical calculations \cite{EFLmt}. In the analysis leading to the possible
discovery of the top quark, the Monte Carlo simulations which are compared with
the data, assume the validity of the Standard Model, and no other processes
beyond it contribute to the sought-for signal. In this note we would like to
point out that, in the context of supersymmetric models, the pair-production of
the lightest top-squark (``stop") may lead to very similar experimental
signatures as the pair-production of top quarks. This fact by itself is not
new, since it is well known that one can always adjust arbitrarily the
parameters of the the Minimal Supersymmetric Standard Model (MSSM) to have a
light top-squark \cite{LightStopOld,BDGGT,Japs,BST}. However, in the context of
the minimal $SU(5)$ supergravity model \cite{Dickreview}, \ie, the simplest
model underlying the MSSM, the constraints from the proton lifetime
\cite{pdecay} force {\em all} the squarks to be heavier than the top quark.
On the other hand, a light top-squark may be the natural consequence of a
one-parameter string-inspired $SU(5)\times U(1)$ supergravity model
\cite{LNZII}, with the dilaton field being the dominant source of supersymmetry
breaking \cite{KL}, {\em and} the electroweak-size Higgs mixing parameter $\mu$
 obtained naturally from supergravity-induced contributions
\cite{KL,BLM,LNZII}.

Our model \cite{LNZII} is a special case of a generic supergravity model with
universal soft supersymmetry breaking, which is described in terms of four
parameters: $m_{1/2},m_0,A,\tan\beta$. In the ``special dilaton" scenario
one has \cite{KL}
\begin{equation}
m_0=\coeff{1}{\sqrt{3}}m_{1/2},\quad A=-m_{1/2},\quad B=2m_0,
\label{relations}
\end{equation}
where $B$ is the soft-supersymmetry-breaking parameter (at the unification
scale) associated with $\mu$. These conditions determine all but one parameter,
taken here to be $m_{1/2}\propto m_{\chi^\pm_1}\propto m_{\tilde g}$. The
requirement of radiative electroweak symmetry breaking,\footnote{For recent
reviews of this general procedure see \eg, Ref.~\cite{reviews}.} which
determines $\mu$ up to a sign, can only be satisfied here for $\mu<0$, in light
of the last condition $B=2m_0$. Moreover, this condition determines $\tan\beta$
as a function of $m_{1/2}$; one finds that $\tan\beta$ must be small:
$\tan\beta\approx1.4$, with little dependence on $m_{1/2}$ \cite{LNZII}. In
what follows we take $m_t=162\GeV$, \ie, the central value of the world-average
fit to $m_t$ ($m_t=162\pm9\GeV$ \cite{EFL}). (Details of the following analysis
will appear elsewhere \cite{LNZIII}.)

\begin{figure}[p]
\vspace{6in}
\includegraphics{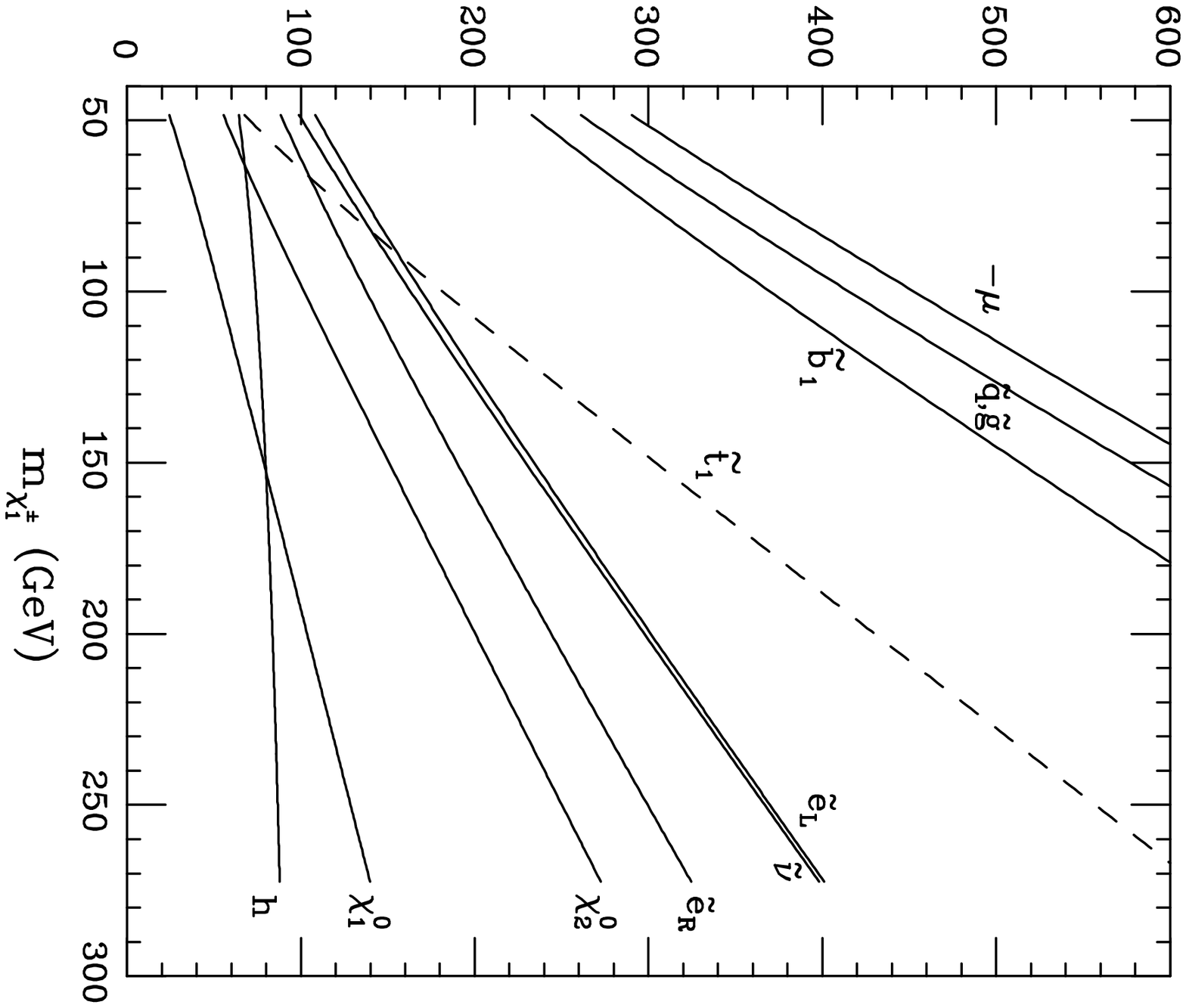}
\vspace{1cm}
\caption{The relevant lighter sparticle masses versus the chargino mass.
The $\tilde t_1$ top-squark mass (with $m_{\tilde t_1}>67\GeV$) is shown by the
dashed line. Note that $m_{\tilde t_1}>m_{\chi^\pm_1}$. Here $m_{\tilde
q}\approx m_{\tilde g}$, with $m_{\tilde q}$ the average first- or
second-generation squark mass. Also, $m_{\chi^0_2}\approx
m_{\chi^\pm_1}\approx2m_{\chi^0_1}$, and $m_A\approx m_H\approx
m_{H^\pm}>400\,{\rm GeV}$.}
\label{Fig1}
\end{figure}

For our present purposes, the main result, \ie, a light top-squark, is a
consequence of the small value of $\tan\beta$. Indeed, the lightest top-squark
mass is given~by
\begin{eqnarray}
m^2_{\tilde t_1}&=&\coeff{1}{2}(m^2_{\tilde t_L}+m^2_{\tilde t_R})
+\coeff{1}{4}M^2_Z\cos2\beta+m^2_t\nonumber\\
&-&\sqrt{\left[\coeff{1}{2}(m^2_{\tilde t_1}-m^2_{\tilde t_R})
+\coeff{1}{12}\cos2\beta(8M^2_W-5M^2_Z)\right]^2+m^2_t(A_t+\mu/\tan\beta)^2}\ ,
\label{stopmass}
\end{eqnarray}
where $m^2_{\tilde t_{L,R}}$ are the running top-squark masses. In the present
case there is a large cancellation between the first term
$\coeff{1}{2}(m^2_{\tilde t_L}+m^2_{\tilde t_R})$ and the last term in the
square root $m^2_t(A_t+\mu/\tan\beta)^2$, which leads to light top-squark
masses, \ie,
\begin{equation}
m^2_{\tilde t_1}\sim \coeff{1}{2}(m^2_{\tilde t_L}+m^2_{\tilde t_R})+m^2_t
-m_t|A_t+\mu/\tan\beta|\ .
\label{cancellation}
\end{equation}
We find $m_{\tilde t_1}\gsim67\GeV$ ({\em c.f.}, the LEP limit $m_{\tilde
t_1}>45\GeV$ \cite{PDG}). (This result has a strong $\tan\beta$ dependence,
\eg, $m_{\tilde t_1}\gsim90\,(120)\GeV$ for $\tan\beta\approx1.5\,(2.0)$, but
here $\tan\beta$ is fixed and cannot be varied at will.) We also find
$m_{\tilde q}\approx m_{\tilde g}\gsim260\GeV$, where $m_{\tilde q}$ is the
average first- or second-generation squark mass. In Fig.~\ref{Fig1} we present
a collection of spectra plots versus the lightest chargino mass
($m_{\chi^\pm_1}$) for the lighter supersymmetric particles. We note in passing
that in this model we find $B(b\to s\gamma)\approx(1-3)\times10^{-4}$,
which is in very good agreement with the present experimental results
\cite{Thorndike}. Also, the relic density of the lightest neutralino satisfies
$\Omega_\chi h^2_0\lsim0.85$, which is in natural agreement with cosmological
observations and includes the possibility of a Universe with a cosmological
constant \cite{CC}.

\begin{table}[t]
\caption{Cross sections at the Tevatron (in pb) for $p\bar p\to\tilde
t_1\bar{\tilde t_1}X$ [5] and $p\bar p\to t\bar tX$~[19]. All masses in GeV.}
\label{Xsections}
\begin{center}
\begin{tabular}{|c|c|c|c|c|c|}\hline
$m_{\tilde t_1}$&70&80&90&100&112\\ \hline
$\sigma(\tilde t_1\bar{\tilde t_1})$&60&30&15&8&4 \\ \hline
\end{tabular}
\begin{tabular}{|c|c|c|c|c|}\hline
$m_t$&120&140&160&180\\ \hline
$\sigma(t\bar t)$&39&17&8&4 \\ \hline
\end{tabular}
\end{center}
\hrule
\end{table}

\begin{figure}[p]
\vspace{6in}
\includegraphics{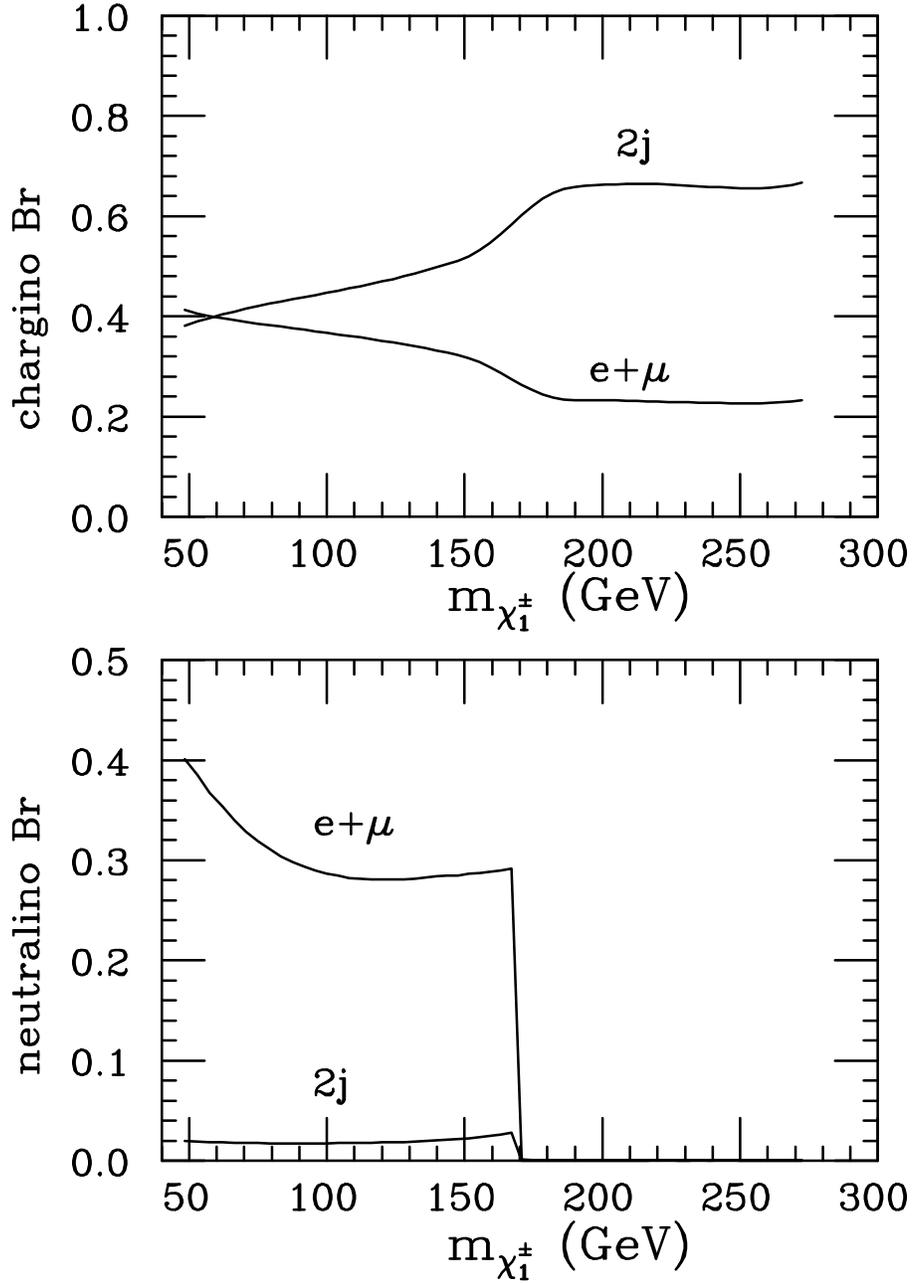}
\vspace{1cm}
\caption{The leptonic and hadronic branching fractions of the chargino
($\chi^\pm_1$) and the neutralino ($\chi^0_2$) (other channels are not shown).
The sudden drop in the leptonic neutralino branching ratio at
$m_{\chi^\pm_1}\approx170\,{\rm GeV}$ corresponds
to the opening of the ``spoiler mode" $\chi^0_2\to\chi^0_1+h$.}
\label{Fig2}
\end{figure}

The cross section for pair-production of the lightest top-squarks
$\sigma(\tilde t_1\bar{\tilde t_1})$ depends solely on $m_{\tilde t_1}$
\cite{BDGGT} and is given for a sampling of values in Table~\ref{Xsections}.
Since in this model $m_{\tilde t_1}>m_{\chi^\pm_1}+m_b$ (see Fig.~\ref{Fig1}),
one gets $B(\tilde t_1\to b\chi^\pm_1)=1$. The charginos then decay
leptonically or hadronically with branching fractions shown in Fig.~\ref{Fig2},
\ie, $B(\chi^\pm_1\to \ell\nu_\ell\chi^0_1)\approx0.4$ ($\ell=e+\mu$) for
$m_{\chi^\pm_1}\lsim65\GeV\leftrightarrow m_{\tilde t_1}\lsim100\GeV$.
The most promising signature for light top-squark detection is through the
dilepton mode \cite{BST}. The number of stop-dileptons is:
\begin{equation}
N^{\tilde t_1\bar{\tilde t_1}}_{2\ell}=\sigma(\tilde t_1\bar{\tilde t_1})\times
[B(\tilde t_1\to b\chi^\pm_1)]^2\times[B(\chi^\pm_1\to \ell\nu_\ell\chi^0_1)]^2
\times{\cal L}\approx0.16\,\sigma(\tilde t_1\bar{\tilde t_1})\times{\cal L}.
\label{stop-dileptons}
\end{equation}
The dilepton mode is also paramount in top-quark searches:
 \begin{equation}
N^{t\bar t}_{2\ell}=\sigma(t\bar t)\times
[B(t\to bW)]^2\times[B(W\to\ell\nu_\ell)]^2\times{\cal L}
\approx0.05\,\sigma(t\bar t)\times{\cal L}.
\label{top-dileptons}
\end{equation}
Here we have taken $B(t\to bW)=1$, although one should account for the
$t\to\tilde t_1\chi^0_1$ mode which is open also for light top-squarks.
Moreover, $p\bar p\to t\bar t X\to \tilde t_1\bar{\tilde t_1}\chi^0_1\chi^0_1X$
is another source of top-squarks, although much suppressed because of the small
branching fraction: we find $B(t\to\tilde t_1\chi^0_1)\lsim10\%$.
Combining Eqs.~(\ref{stop-dileptons},\ref{top-dileptons}) we obtain
\begin{equation}
{N^{\tilde t_1\bar{\tilde t_1}}_{2\ell}\over N^{t\bar t}_{2\ell}}
\approx 3.2{\sigma(\tilde t_1\bar{\tilde t_1})\over \sigma(t\bar t)}\ .
\label{ratio}
\end{equation}
This ratio should open the eyes of experimenters because the number of observed
dilepton events depends strongly on the experimental biases. This ratio
(\ref{ratio}) indicates that for sufficiently light top-squarks there
may be a significant number of dilepton events of non--top-quark origin,
if the experimental acceptances are tuned accordingly.

Perhaps the most important distinction between top-dileptons and
stop-dileptons is their $p_T$ distribution: the (harder) top-dileptons come
from the two-body decay of the $W$ boson, whereas the (softer) stop-dileptons
come from the (usually) three-body decay of the chargino with masses (in
this case) below $m_W$. Therefore, the top-dilepton data sample is essentially
distinct from the stop-dilepton sample. Such distinction is well quantified by
the ``bigness" ($B$) parameter $B=|p_T(\ell^+)|+|p_T(\ell'^-)|+|\met|$ of
Ref.~\cite{BST}. Another distinction between the two sources of dileptons are
the $b$-jets, which are probably softer in the decay $\tilde t_1\to
b\chi^\pm_1$ (for light top-squarks) compared to those from $t\to bW$. The
above discussion suggests that the CDF top-dilepton data sample should be
carefully studied to see if softer stop-dileptons are present: an important new
lower bound on the top-squark mass may follow. However, detailed simulations of
the stop-dilepton signal and a re-analysis of the top-dilepton data are
required before drawing more concrete conclusions.

We also note that in the $\ell$+jets channel, the ratio analogous to
Eq.~(\ref{ratio}) is ${N^{\tilde t_1\bar{\tilde t_1}}_{\ell+{\rm jets}}/
N^{t\bar t}_{\ell+{\rm jets}}}
\approx {\sigma(\tilde t_1\bar{\tilde t_1})/ \sigma(t\bar t)}$, since
$B(W\to 2j)\cdot B(W\to \ell)=(2/3)(2/9)\approx B(\chi^\pm_1\to 2j)\cdot
B(\chi^\pm_1\to\ell)$ (see Fig.~\ref{Fig2}). In this case, the top-squark
$\ell$+jets events still have softer $b$-jets and a softer lepton.

\begin{table}[t]
\caption{Upper limits on sparticle masses which follow from $m_{\tilde
t_1}<100\,{\rm GeV}$, such that $\tilde t_1$ may be relevant in top-quark
searches. All masses in GeV.}
\label{limits}
\begin{center}
\begin{tabular}{|c|c|c|c|c|c|c|c|c|c|}\hline
$\chi^\pm_1$&$\chi^0_1$&$\chi^0_2$&$h$&$\tilde e_R$
&$\tilde\nu$&$\tilde e_L$&$\tilde t_1$&$\tilde b_1$&$\tilde q,\tilde g$\\
\hline
65&35&70&70&108&120&130&100&275&310\\ \hline
\end{tabular}
\end{center}
\hrule
\end{table}

\begin{figure}[p]
\vspace{6in}
\includegraphics{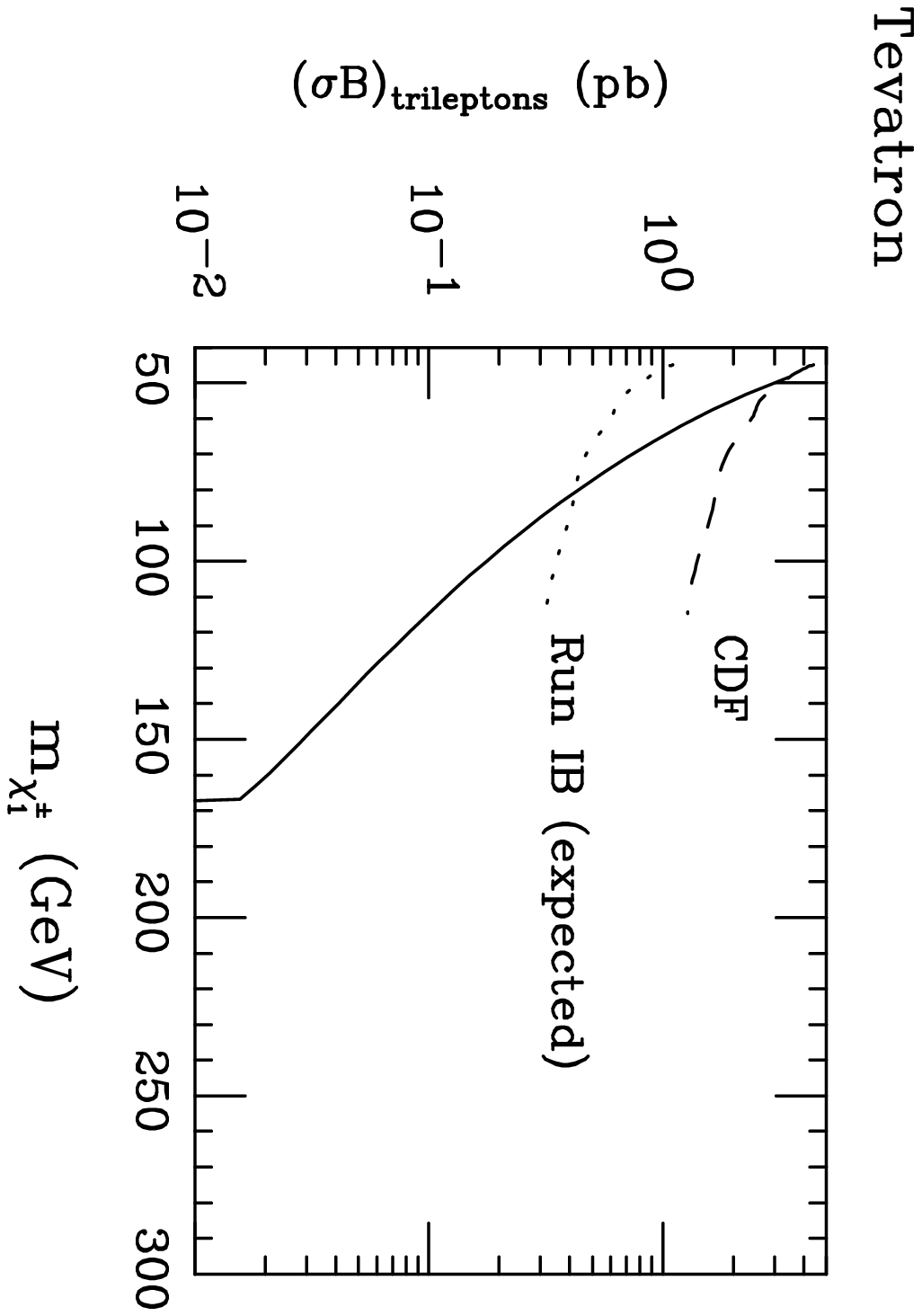}
\vspace{1cm}
\caption{The rate for trilepton events at the Tevatron. The present CDF limit
is indicated. The dotted line indicates the expected sensitivity at the end
of Run IB ($\sim100\,{\rm pb}^{-1}$) equivalent to a reach
$m_{\chi^\pm_1}<80\,{\rm GeV}$.}
\label{Fig3}
\end{figure}

The light top-squarks which may be relevant for the top-quark and top-squark
searches at the Tevatron (\ie, $m_{\tilde t_1}\lsim100\GeV$) entail a light
supersymmetric spectrum, as can be seen from Fig.~\ref{Fig1}. For $m_{\tilde
t_1}<100\GeV$, we get the corresponding upper limits shown in
Table~\ref{limits}. We now explore the possibilities for direct detection of
these light sparticles at the Tevatron and LEPII.
\begin{itemize}
\item {\em Tevatron}. One could detect these light sparticles in three ways:
\begin{itemize}
\item The trilepton signal in $p\bar p\to\chi^\pm_1\chi^0_2X$ is the most
promising avenue for detection of weakly interacting sparticles at the Tevatron
\cite{EHNS,trileptons}, as evidenced in the context of $SU(5)\times U(1)$
supergravity in Ref.~\cite{LNWZ}. The leptonic chargino and neutralino
branching fractions are given in Fig.~\ref{Fig2}, and the trilepton rate at the
1.8~TeV Tevatron is given in Fig.~\ref{Fig3}, where we indicate by a dashed
line the present CDF upper limit \cite{Kato} and by a dotted line the expected
reach by the end of Run IB (with $\sim100\ipb$ of accumulated data). This reach
corresponds to $m_{\chi^\pm_1}\lsim80\GeV\leftrightarrow m_{\tilde
t_1}\lsim130\GeV$. Therefore, the light sector of this model -- that relevant
to top-quark searches -- could be definitively falsified in the next few
months.
\item Direct $\tilde t_1$ pair production at the Tevatron has been shown
recently \cite{BST} to be sensitive to $m_{\tilde t_1}\lsim100\GeV$ by the
end of Run IB, provided the chargino leptonic branching fraction is taken
to be $\sim20\%$. For the chargino branching fractions in our model
($\sim40\%$, see Fig.~\ref{Fig2}) the reach through the stop-dilepton channel
is extended to $m_{\tilde t_1}\lsim130\GeV$.
\item The standard squark-gluino searches may also be able to reach
up to $m_{\tilde q}\approx m_{\tilde g}\approx310\GeV$ with the Run IB data.
\end{itemize}
\item {\em LEPII}. One could detect these light sparticles in three ways:
\begin{itemize}
\item Charginos would be readily pair-produced, and best detected
through the ``mixed" mode (\ie, $\ell+2j$). For
$m_{\chi^\pm_1}\lsim65\,(80)\GeV$, we find $(\sigma\times B)_{\rm
mixed}\gsim0.34\,(0.27)\pb$, which is much larger than
the estimated $5\sigma$ sensitivity at $100\ipb$, \ie, $0.12\pb$
\cite{Easpects}.
\item The lightest Higgs boson should be easily detectable through the standard
process $e^+e^-\to Z^*\to Zh$. For $m_h\lsim70\GeV$ (from Table~\ref{limits}),
we find a cross section in excess of $0.92\pb$, which is much larger than the
expected sensitivity limit of $0.2\pb$ for a $3\sigma$ effect at $500\ipb$
\cite{Sopczak}. In fact, a $0.92\pb$ signal corresponds to a significance of
$6.2\sigma$ at $100\ipb$.
\item The light top-squark may also be produced directly $e^+e^-\to\tilde
t_1\bar{\tilde t_1}$ via $s$-channel $\gamma,Z$ exchange, and be probed up to
$m_{\tilde t_1}\approx\sqrt{s}/2\approx100\GeV$.
\end{itemize}
\end{itemize}

In summary, we have discussed the prediction of a light top-squark in a
string-inspired one-parameter $SU(5)\times U(1)$ supergravity model. This
sparticle ($\tilde t_1$) may be readily pair-produced at the Tevatron and, if
$m_{\tilde t_1}\lsim130\GeV$, even be observed with the present run accumulated
data. Top-squark production may also be an important source of sought-for
top-quark signatures in the dilepton and $\ell$+jets channels. Therefore, a
re-analysis of the top data sample in the presence of a possibly light
top-squark appears necessary before definitive statements concerning the
discovery of the top quark can be made. Another prediction of this model is
a direct link between the light top-squark and a light supersymmetric spectrum,
which can certainly be searched for at the Tevatron through trilepton and
squark-gluino searches, and at LEPII through direct $\tilde t_1$
pair-production (for $m_{\tilde t_1}\lsim100\GeV$) and via chargino and
Higgs-boson searches.

\section*{Acknowledgments}
This work has been supported in part by DOE grant DE-FG05-91-ER-40633. We would
like to thank Teruki Kamon, Peter McIntyre, and James White for useful
discussions.

\newpage

\clearpage

\clearpage

\clearpage

\end{document}